\documentclass[fleqn,usenatbib]{mnras}

\usepackage{newtxtext,newtxmath}

\usepackage[T1]{fontenc}
\usepackage{color}
\usepackage{lineno}
\usepackage{diagbox}

\DeclareRobustCommand{\VAN}[3]{#2}
\let\VANthebibliography\thebibliography
\def\thebibliography{\DeclareRobustCommand{\VAN}[3]{##3}\VANthebibliography}

\usepackage{graphicx}	
\usepackage{amsmath}	

\title[Microlensing bias on SLGWs]{Microlensing bias on the detection of strong lensing gravitational wave}

\author[X. Shan et al.]{
Xikai Shan,$^{1,2}$
Xuechun Chen,$^{1,2}$
Bin Hu$^{1,2}$\thanks{E-mail: bhu@bnu.edu.cn}
and Guoliang Li$^{3}$
\\
$^{1}$Institute for Frontier in Astronomy and Astrophysics, Beijing Normal University, Beijing, 102206, China\\
$^{2}$Department of Astronomy, Beijing Normal University, Beijing, 100875, China\\
$^{3}$Purple Mountain Observatory, Chinese Academy of Sciences, Nanjing, Jiangsu, 210023, China}

\date{Accepted XXX. Received YYY; in original form ZZZ}

\pubyear{2023}

\begin{document}
\label{firstpage}
\pagerange{\pageref{firstpage}--\pageref{lastpage}}
\maketitle

\begin{abstract}
Identifying strong lensing gravitational wave (SLGW) events is of utmost importance in astrophysics as we approach the historic first detection of SLGW amidst the growing number of gravitational wave (GW) events.
Currently, one crucial method for identifying SLGW signals involves assessing the overlap of parameters between two GWs.
However, the distribution of discrete matter, such as stars and sub-halos, within the strong lensing galaxy can imprint a wave optical (WO) effect on the SLGW waveform.
These frequency dependent imprints introduce biases in parameter estimation and impact SLGW identification.
In this study, we assess the influence of the stellar microlensing field embedded in a strong lensing galaxy.
Our finding demonstrate that the WO effect reduces the detection efficiency of SLGW by $5\%\sim 50\%$ for various false alarm probabilities per pair (${\rm FAP}_{\rm per~pair}$). 
Specifically, at an ${\rm FAP}_{\rm per~pair}$ of $10^{-5}$, the detection efficiency decreases from $\sim 10\%$ to $\sim 5\%$.
Consequently, the presence of the microlensing field can result in missing half of the strong lensing candidates.
Additionally, the microlensing WO effect introduces a noticeable bias in intrinsic parameters, particularly for chirp mass and mass ratio.
However, it has tiny influence on extrinsic parameters. 
Considering all parameters, $\sim 30\%$ of events exhibit a $1\sigma$ parameter bias, $\sim 12\%$ exhibit a $2\sigma$ parameter bias, and $\sim 5\%$ exhibit a $3\sigma$ parameter bias.
\end{abstract}




\section{Introduction}
When light passes near a massive object, its path is bent by gravity, resulting in the formation of multiple images known as the strong gravitational lensing effect. 
However, Within the lens galaxy, the presence of microlens fields leads to observed characteristics of strong lensing images that go beyond what is predicted by the standard strong lens model. 
This phenomenon is referred to as the microlensing effect.

In 1991, \citet{1991ApJ...373..354K} made a significant discovery.
They found that while a simple smooth lens galaxy model could accurately predict the positions of the lensed images, it failed to account for their magnifications, indicating the existence of substructures within lens galaxies. 
Subsequent studies by researchers such as~\citet{Mao:1997ek, Metcalf:2001ap, Chiba:2001wk, Dalal:2001fq, Metcalf:2001es, Keeton:2002qt, Kochanek:2003zc, Bradac:2003hy} further explored the impact of these substructures and observed anomalies in the flux ratios of strongly lensed image pairs caused by subhalos in the lens galaxy.
These findings emphasized the importance of considering substructures/microlens field in the study of strong lensing.
Similarly, different types of strong lensing gravitational wave (SLGW) signals are influenced by various substructures, potentially leading to phenomena resembling flux ratio anomalies.

However, the long-wavelength nature of gravitational waves (GWs) renders the gravitational lensing effect of substructures, particularly the stellar microlens field examined in this study, unable to be accurately calculated using geometric optical approximations.
Instead, the interference and diffraction effects become crucial due to the shorter time delay between multiple micro-images produced by microlenses compared to the period of gravitational waves.
This effect is known as the WO effect.
Precisely speaking, when lens mass $\mathrm{M}_\mathrm{L}$ and GW frequency $f$ satisfy the following relationship, the WO effect is noticeable~\citep{Takahashi:2003ix}.
\begin{equation}
\label{eq:geo_wo}
\mathrm{M}_\mathrm{L} \lesssim 10^{5} M_{\odot}\left(\frac{f}{\mathrm{Hz}}\right)^{-1} \,
\end{equation}
where $M_{\odot}$ is the solar mass.

The most notable feature of the WO effect is that it can produce a frequency-dependent fluctuation in the GW's waveform.
Numerous pioneering works have focused on exploring this aspect.
\citet{Cheung:2020okf} and \citet{Yeung:2021roe} studied the WO effect due to a point mass embedded in different strong lensing images (Type I and Type II).
\citet{Meena:2019ate} analyzed the influence of a point mass lens on GW's parameter estimation for different strong lensing images and found that it can lead to detectable differences in the signal. 
Some works also focused on the microlensing field embedded in the strong lensing galaxy~\citep[e.g.,][]{2019Diego,2021Anuj, Meena:2022unp, Shan:2022xfx, Shan:2023ngi, Savastano:2023spl,Mishra:2023ddt}, revealing its distinct characteristics compared to an isolated point-mass lens.

As the number of gravitational wave (GW) events rapidly increases~\citep{LIGOScientific:2018mvr, abbott2021gwtc2, LIGOScientific:2021djp}, the search for SLGW has become a prominent area of research in GW astronomy.
In recent years, various methods have been developed to search for SLGWs, including assessing the degree of parameter overlap between image pairs~\citep{Haris:2018vmn, Dai:2020tpj, Lo:2021nae, Gao:2023uxi}, employing machine learning techniques to identify similar time-frequency structures~\citep{Kim:2020xkm, Goyal:2021hxv}, searching for saddle point image characteristics when GW's high order modes are significant~\citep{Dai:2017huk, Wang:2021kzt, Janquart:2021nus}, and searching for microlensing field imprints on SLGW~\citep{Shan:2023ngi}.
For further in-depth information, one may refer to the latest review~\citep{Liao:2022gde}.

However, the first three studies did not take into account the potential bias caused by the microlensing WO effect.
In this paper, we aim to forecast the influence of the WO effects arising from microlensing fields on the overlapping method and investigate the microlensing bias on GW parameter estimation from a statistical perspective.
In this analysis, we utilize two Cosmic Explorer (CE)~\citep{Evans:2016mbw} detectors located at Livingston and Hanford.

This paper is organized as follows. 
Section~\ref{sec:met} provides an overview of the basic theory of GW lensing and outlines the methodology employed for simulating SLGWs.
In Section~\ref{sec:res}, we present the findings regarding the microlensing effect on the detection efficiency of SLGWs and present the results of parameter estimation bias.
Finally, we do summaries and discussions in Section~\ref{sec:sum_dis}.

\section{Methodology}
\label{sec:met}
\subsection{Strong lensing and microlensing}
The strong gravitational lensing equation of GWs generated by stellar-mass binary black holes is the same as the optical rays:
\begin{equation}
\label{eq:leq}
\mathbf{y}=\mathbf{x}-\boldsymbol{\alpha}(\mathbf{x}) \\,
\end{equation}
where $\mathbf{y}$ and $\mathbf{x}$ are source and image angle positions of GW.
$\boldsymbol{\alpha}(\mathbf{x})$ is the rescaled deflection angle:
\begin{equation}
\label{eq:def_ang}
\boldsymbol{\alpha}(\mathbf{x})=\nabla \psi(\mathbf{x}) \\.
\end{equation}
$\psi(\mathbf{x})$ is the rescaled two-dimensional lens potential which defined as
\begin{equation}
\label{eq:2dP}
\psi(\mathbf{x})=\frac{1}{\pi} \int d^{2} \mathbf{x}^{\prime} \kappa\left(\mathbf{x}^{\prime}\right) \ln \left|\mathbf{x}-\mathbf{x}^{\prime}\right| \\,
\end{equation}
where $\kappa\left(\mathbf{x}\right)$ is the dimensionless surface-mass density.
One can find that given the distribution of $\kappa\left(\mathbf{x}\right)$, the deflection angle $\boldsymbol{\alpha}(\mathbf{x})$ can be determined by using Eq.~(\ref{eq:def_ang}) and Eq.~(\ref{eq:2dP}).
However, to get the image positions for a specified source position, one needs to solve the lens equation Eq.~(\ref{eq:leq}).
In this paper, we use the public package \texttt{Lenstronomy}~\citep{2018PDU....22..189B, 2021JOSS....6.3283B} to complete this complicated calculation.

After solving the lens equation, one can get the time delay between multiple SLGWs.
\begin{equation}
\label{eq;t-d}
\Delta t(\mathbf{y})=\frac{\left(1+z_{\mathrm{L}}\right)}{c} \frac{D_{\mathrm{s}} D_{\mathrm{L}}}{D_{\mathrm{Ls}}}\left[\phi\left(\mathbf{x}_{1}, \mathbf{y}\right)-\phi\left(\mathbf{x}_{2}, \mathbf{y}\right)\right] \\,
\end{equation}
where $z_\mathrm{L}$ is the lens redshift, $c$ is the light speed.
$D_\mathrm{s}$, $D_\mathrm{L}$, and $D_\mathrm{Ls}$ are the angular diameter distances for observer-source, observer-lens, and lens-source.
$\mathbf{x}_1$ and $\mathbf{x}_2$ are the angular positions of signal $1$ and signal $2$ in the lens plane.
$\phi\left(\mathbf{x}, \mathbf{y}\right)$ is the Fermat potential
\begin{equation}
\phi(\mathbf{x}, \mathbf{y})=\frac{1}{2}(\mathbf{x}-\mathbf{y})^{2}-\psi(\mathbf{x}) \\.
\end{equation}
The magnification for individual SLGW is
\begin{equation}
\mu\left(\mathbf{x}\right) = \frac{1}{(1-\kappa\left(\mathbf{x}\right))^2 - \gamma\left(\mathbf{x}\right)^2} \\.
\end{equation}
$\gamma\left(\mathbf{x}\right)$ is the gravitational lensing shear which is defined as:
\begin{equation}
\gamma\left(\mathbf{x}\right) = \sqrt{\gamma_{1}\left(\mathbf{x}\right)^{2}+\gamma_{2}\left(\mathbf{x}\right)^{2}} \\,
\end{equation}
where
\begin{equation}
\begin{aligned}
\gamma_{1}\left(\mathbf{x}\right) & =\frac{1}{2}\left(\frac{\partial{^2\psi\left(\mathbf{x}\right)}}{\partial{x_1}\partial{x_1}}-\frac{\partial^2{\psi\left(\mathbf{x}\right)}}{\partial{x_2}\partial{x_2}}\right) 
\\ 
\gamma_{2}\left(\mathbf{x}\right) & =\frac{\partial{^2\psi\left(\mathbf{x}\right)}}{\partial{x_1}\partial{x_2}}
\end{aligned}
\end{equation}

In real lens galaxies, as we mentioned earlier, in addition to the smooth mass distribution, there are many sub-structures, such as dark matter sub-halos and stellar fields.
Therefore, we also need to consider these effects on SLGWs.
In this paper, we only evaluate the influence of stellar microlensing fields and use the point mass lens model to describe the microlens.

Since the stellar mass and the frequency of GW considered here ($10\sim1000~\mathrm{Hz}$) satisfy Eq.~(\ref{eq:geo_wo}), we need to evaluate the influence of the microlensing fields on SLGWs using the WO method.
The WO effect of SLGW can be quantified by the diffraction integral~\citep{schneider1992gravitational,10.1143/PTPS.133.137,Takahashi:2003ix}
\begin{equation}
\label{eq:DiffInter}
F(\omega)=\frac{2 G \mathrm{M}_\mathrm{L}\left(1+z_\mathrm{L}\right) \omega}{\pi c^{3} i} \int_{-\infty}^{\infty} d^{2} x^\prime \exp \left[i \omega t(\boldsymbol{x^\prime})\right] ,
\end{equation}
where $F(\omega)$ is the magnification factor, $\omega$ is the circle frequency of GW.
The microlensing time-delay function is
\begin{equation}
\begin{split}
\label{eq:TimeDelay}
t(\boldsymbol{x^\prime}) = &\underbrace{\frac{k}{2}\left((1-\kappa+\gamma) x_{1}^{\prime 2}+(1-\kappa-\gamma) x_{2}^{\prime 2}\right)}_{t_\text{s}(\kappa,\gamma,\boldsymbol{x}^\prime)} \\ 
& - \underbrace{\left[\frac{k}{2}\sum_{i}^{N} \ln \left(\boldsymbol{x}^{\prime i}-\boldsymbol{x^\prime}\right)^{2} + k\phi_{-}(\boldsymbol{x^\prime})\right]}_{t_\text{m}(\boldsymbol{x^\prime},\boldsymbol{x}^{\prime i})}
\end{split}
\end{equation}
where $k=4 G \text{M}_\mathrm{L}(1+z_\mathrm{L})/c^3$, $\boldsymbol{x^{\prime i}}$ is the $i$th microlens coordinate in the lens plane, $t_\text{s}(\kappa,\gamma,\boldsymbol{x}^\prime)$ and $t_\text{m}(\boldsymbol{x^\prime},\boldsymbol{x}^{\prime i})$ are the time delays induced by the strong lens and microlens potentials, respectively.
$\phi_{-}(\boldsymbol{x^\prime})$ is the negative mass sheet's contribution to keeping the total convergence $\kappa$ unchanged when adding microlenses~\citep{Wambsganss1990, 2021xuechunchen, Zheng:2022vfq}.
It is important to highlight that the coordinate systems used for $\boldsymbol{x^\prime}$ and $\boldsymbol{x}$ mentioned earlier are not the same.
Specifically, $\boldsymbol{x^\prime}$ and $\boldsymbol{x}$ have been rescaled with the Einstein radius of the microlens and the strong lens, respectively.

\subsection{Mock data simulation}
\label{sec:moc_dat}
To evaluate the microlensing WO effect on the SLGWs, one needs to simulate a mock data set.
Here, we use the simulation process following~\cite{Haris:2018vmn,Xu:2021bfn,Shan:2023ngi}.

The first step is to construct binary black holes (BBHs) distribution.
In this paper, we use an analytical BBH merge rate model:
\begin{equation}
\label{eq:mer_rat}
\frac{d \dot{N}(z)}{d z} \equiv \frac{\mathcal{R}(z)}{(1+z)} \frac{d V(z)}{d z} \\,
\end{equation}
where $\mathcal{R}(z)$ is the source-frame merger rate of BBH.
The factor $1+z$ converts the merger rate from the source to the observer frame.
Here, we follow the strategy in refs.~\citep{Haris:2018vmn,Xu:2021bfn,Shan:2023ngi}, assuming that the merger rate is related to the star formation rate (SFR)
\begin{equation}
\mathcal{R}(z)=C \int_{\Delta t_{\min }}^{t_{H}(z)} \dot{\rho}_{*}\left(z_{bf}\right) P(\Delta t) d \Delta t \\,
\end{equation}
where $z_{bf}$ is the binary formation redshift.
$\Delta t$ is the delay time between binary formation and BBH merger.
Coefficient $C$ is a normalization factor, which guarantees $\mathcal{R}(z=0)=64.9^{+75.5}_{-33.6}\mathrm{Gpc}^{-3}\mathrm{yr}^{-1}$~\citep{LIGOScientific:2018jsj}.
$\dot{\rho}_{*}\left(z_{bf}\right)$ is the SFR which describes the formation number of stars per unit volume and time.
We use the parameterization form introduced in ~\citet{Madau:2014bja}.
\begin{equation}
\dot{\rho}_{*}(z)=0.015 \frac{(1+z)^{2.7}}{1+[(1+z) / 2.9]^{5.6}} M_{\odot} \mathrm{Mpc}^{-3} \mathrm{yr}^{-1}
\end{equation}
The delay-time distribution form is $P(\Delta t) \propto \frac{1}{\Delta t}$, and we assume the minimum delay-time $\Delta t_{\min}$ is $50\mathrm{Myr}$.

After integrating $z$ in Eq.~(\ref{eq:mer_rat}), one can find that the total number of BBH merges in a year is $\sim 1.3 \times 10^{5}$.
So we randomly sample $\sim 1.3 \times 10^{5}$ BBH redshifts from Eq.~(\ref{eq:mer_rat}) using the Monte Carlo method.
Now, we finished the redshift assignment in the BBH mock data set. 

For the GW events picked above, we randomly assign BBH masses ($m_1$, $m_2$), polarization angle (POL), right ascension angle (RA), declination (DEC), inclination angle ($\iota$), merger time ($t_c$), and spins ($a_1$, $a_2$) from the following distributions:
\begin{itemize}
\item [a)]
$p(m_1)\propto m_1^{-0.4}$, $m_1 \in [5~\mathrm{M}_\odot, 41.6~\mathrm{M}_\odot]$~\citep{LIGOScientific:2018jsj}.
\item [b)]
$p(m_2)\propto \mathrm{U}(5~\mathrm{M}_\odot, m_1)$, where $\mathrm{U}(a, b)$ stands for a uniform distribution from $a$ to $b$.
\item [c)]
$p(\iota)\propto \sin(\iota)$, $\iota \in  [0, \pi]$.
\item [d)]
$p(\mathrm{POL})\propto \mathrm{U}(0,\pi)$.
\item [e)]
$p(\mathrm{RA})\propto \mathrm{U}(0,2\pi)$.
\item [f)]
$p(\mathrm{DEC})\propto \cos(\mathrm{DEC})$, $\mathrm{DEC} \in [-\pi/2, \pi/2]$.
\item [g)]
$p(t_c)\propto \mathrm{U}(t_\mathrm{min}, t_\mathrm{max})$, where $t_\mathrm{min}$ and $t_\mathrm{max}$ are the minimum and maximum merger times used in the simulation. Here, we set $t_\mathrm{max}-t_\mathrm{min}=1\mathrm{yr}$.
\item [h)]
$p(a_1)\propto \mathrm{U}(0, 0.99)$.
\item [i)]
$p(a_2)\propto \mathrm{U}(0, 0.99)$.
\end{itemize}
Then, we adopt the \texttt{IMRPhenomPv2}~\citep{Hannam:2013oca, 2020ascl.soft12021L} waveform model to generate the simulated strain data using \texttt{PyCBC}~\citep{alex_nitz_2022_6324278} .
Here, we use two CE detectors located at Livingston and Hanford.
So far, we have completed the construction of the GW data set.

The second step is to build a lensed catalog from the above simulated unlensed GW data set according to some lensing probability.
\citet{Haris:2018vmn} showed that the multiple-image optical depth of strong lensing is
\begin{equation}
\tau(z_s)=4.17 \times 10^{-6}\left(\frac{D_{\mathrm{s}}^{\mathrm{c}}}{\mathrm{Gpc}}\right)^{3} \\,
\end{equation}
where $D_{\mathrm{s}}^{\mathrm{c}}$ is the comoving distance of the GW source.
The optical depth determines the probability of GW being lensed and having multiple images at redshift $z_s$.
Therefore, we can pick up SLGWs by comparing the optical depth $\tau(z_s)$ with a random number $\xi$ uniformly distributed between $0$ and $1$.
If $\tau(z_s)$ is greater than $\xi$, this GW event would be a SLGW event.
In this step, we have constructed $\sim 100\times n$ SLGWs out of $1.3\times 10^5$ GWs, where $n$ is the image number in a SLGW system.

The third step is to assign strong lens information to the selected SLGW events, that is, strong lensing convergence and shear ($\kappa$ and $\gamma$) at the SLGW image point.
We assumed the singular isothermal ellipsoid (SIE) lens model~\citep{1994A&A...284..285K} for each SLGW event to obtain the above physical quantities.
The velocity dispersion $\sigma_v$ and the axis ratio $q$ used in the SIE lens model are generated from the SDSS galaxy population distribution~\citep{2015ApJ...811...20C, Haris:2018vmn, Wierda:2021upe}. 
In detail, we pick up a parameter $a$ from the following distribution
\begin{equation}
p(x)=x^{\alpha-1} \exp \left(-x^{\beta}\right) \frac{\beta}{\Gamma(\alpha / \beta)} \\,
\end{equation}
where $\alpha=2.32$, $\beta=2.67$, and $\sigma_v$ is equal to $161 \mathrm{km s^{-1}} \times a$.
Then we draw a parameter $b$ from the following distribution
\begin{equation}
g(x)=\frac{x}{s^{2}} \exp \left(-\frac{x^{2}}{2 s^{2}}\right), \quad 0<x<\infty \\,
\end{equation}
where 
\begin{equation}
s=0.38+0.09177 a \\.
\end{equation}
until $b < 0.8$, we set the axis ratio $q = 1 - b$.
For the lens redshift, we draw a sample $r$ from
\begin{equation}
p(x)=30 x^{2}(1-x)^{2}, \quad 0<x<1 \\,
\end{equation}
and the lens comoving distance $D^c_{z_\mathrm{L}} = rD^c_{z_\mathrm{s}}$.
One can get the lens redshift using the comoving distance-redshift relationship.

For the angular position of SLGW, we draw $y_1$ and $y_2$ from the following uniform distributions:
\begin{equation}
\begin{array}{l}y_{1} \in \mathrm{U}\left(0, \sqrt{\frac{q}{1-q^{2}}} \operatorname{arccosh}\left[\frac{1}{q}\right]\right), \\ y_{2} \in \left\{\begin{array}{ll} \mathrm{U}\left(0, \sqrt{\frac{q}{1-q^{2}}} \arccos [q]\right), & \text { if } q>q_{0} \\ \mathrm{U}\left(0, \sqrt{\frac{1}{q}}-\sqrt{\frac{q}{1-q^{2}}} \arccos [q]\right), & \text { if } q<q_{0}\end{array}\right.\end{array} \\,
\end{equation}
where $q_0=0.3942$.
Since we are simulating the multiplied images, we use \texttt{Lenstronomy} to solve the lens equation for picked $y_1$ and $y_2$ until having multiple solutions.
By solving the lens equation, we can obtain the required strong lens parameters ($\kappa$, $\gamma$).

The fourth step is to assign microlens field convergences $\kappa_*$ to SLGWs.
Here, we use three different microlens field densities $f_* = 0.4, 0.6, 0.8$~\citep{2006ApJ...653.1391D} ($\kappa_* = f_* \times \kappa$) for each SLGW.

\begin{figure}
\centering
\vspace{0.1em}\includegraphics[width=\columnwidth]{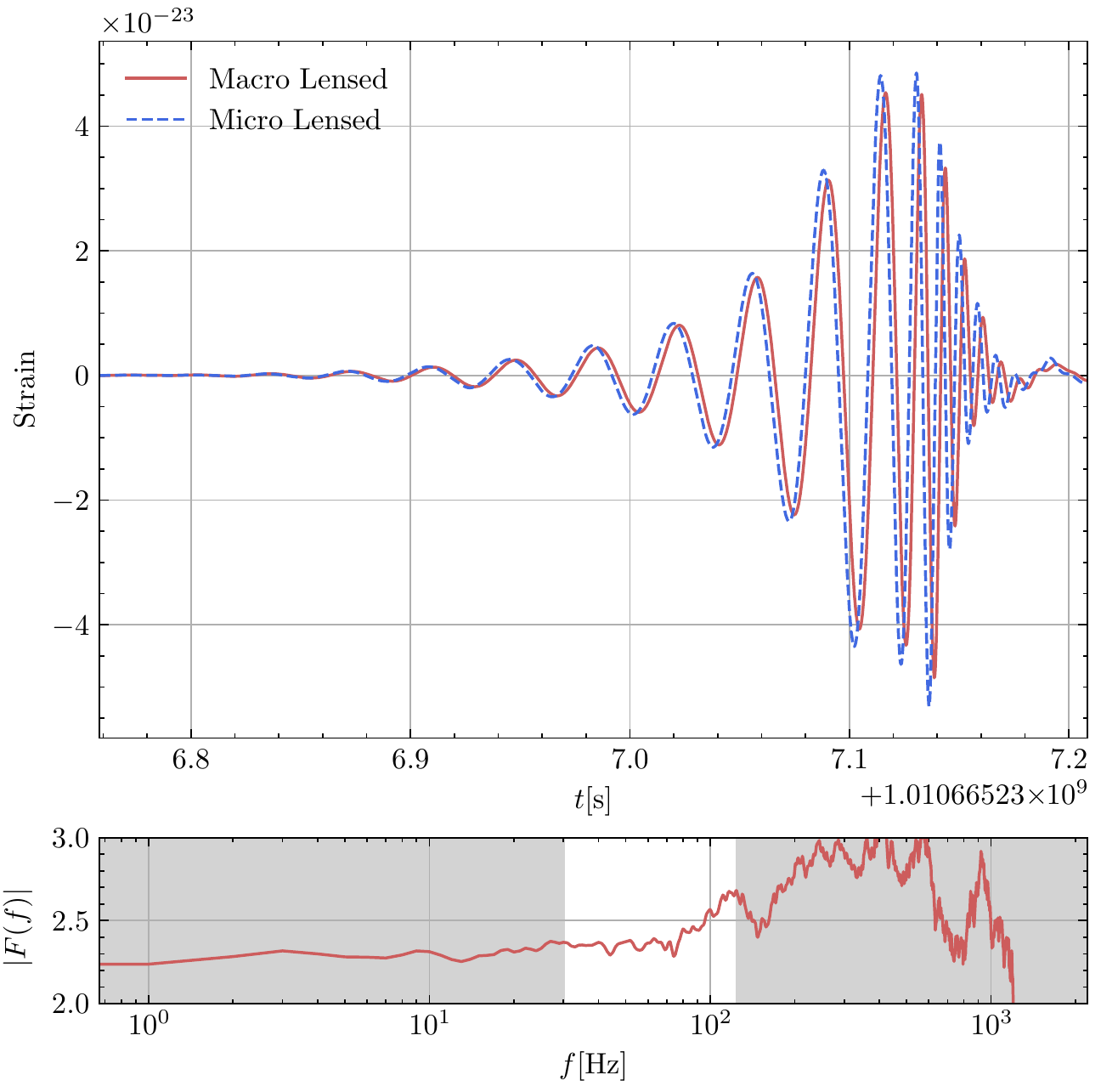}
\caption{This figure shows the impact of microlensing on the SLGW waveform.
The top panel visually represents the time-domain waveform, where the red and blue curves correspond to the SLGW waveform without and with the microlensing effect, respectively. 
The lower panel shows the frequency-domain magnification factor $|F(f)|$ associated with the blue curve in the top panel.
The white window region represents the frequency range of this SLGW event.
}
\label{fig:waveform}
\end{figure}

The fifth and last step is to add the WO effect to each SLGW.
Here, we use the method proposed in~\citet{Shan:2022xfx} to evaluate the diffraction integral Eq.~(\ref{eq:DiffInter}).
The waveform of SLGW with microlensing effect is:
\begin{equation}
\label{eq:fdwf}
\tilde{h}^{L}_{+,\times}(f) = F(f)\tilde{h}_{+,\times}(f) \\,
\end{equation}
where $\tilde{h}_{+,\times}(f)$ and $\tilde{h}^{L}_{+,\times}(f)$ are the unlensed and lensed GW waveforms, respectively.
In Fig.~\ref{fig:waveform}, we present a comparison of the microlensed SLGW waveform with the SLGW waveform unaffected by microlensing in the time domain.
In top panel, the red and blue curves correspond to the SLGW waveform without and with the microlensing effect, respectively.
Comparing these two curves, there is a distinguishable phase shift in the waveform. 
The lower panel showcases the frequency-domain magnification factor $|F(f)|$ associated with the blue curve depicted in the top panel.
The white window region represents the frequency range of this SLGW event.

\section{Result}
\label{sec:res}
In this section, we show the influence of the microlensing field on the identification and parameter estimation of SLGWs.
Through the numerical simulation in Section~\ref{sec:moc_dat}, we find that two CE detectors can observe $178$ SLGW events a year, including $27$ single images (where only one image is visible in a lens system), $68 \times 2$ double images, $1\times3$ triple images (one of the quadruple images are missing), and $3\times4$ quadruple images.
These findings align with the results reported by~\citet{Yang:2019jhw, Yang_2021, Xu:2021bfn}.
Here, we assume the GW event can be detected if its matched filter SNR $\geq 12$~\citep{KAGRA:2013rdx}.

Before presenting the results, we provide a brief overview of the ``overlapping'' method proposed by~\citet{Haris:2018vmn}.
This method utilizes a Bayes factor to distinguish SLGWs from the unlensed GWs.
\begin{equation}
\mathcal{B}_{\mathrm{U}}^{\mathrm{L}}:=\int d \boldsymbol{\theta} \frac{P\left(\boldsymbol{\theta} \mid d_{1}\right) P\left(\boldsymbol{\theta} \mid d_{2}\right)}{P(\boldsymbol{\theta})} \\,
\end{equation}
where $\theta$ represents the parameter of GW, $d_1$ and $d_2$ denotes the strain data for event $1$ and event $2$, respectively.
$P(\boldsymbol{\theta})$ corresponds to the prior distribution, and $P\left(\boldsymbol{\theta} \mid d_{1} (d_{2})\right)$ represents the posterior distribution, which is estimated by using a public package \texttt{Bilby}~\citep{Ashton:2018jfp}.
Here, we only use three parameters, RA, DEC, and $\mathcal{M}_z$ (chirp mass at the detector frame), when calculating the Bayes factor.
The Bayes factor quantifies the similarity between event $1$ and event $2$.
Since the multiple images of SLGWs originate from the same source, the Bayes factor between these events is statistically higher compared to two randomly matched events.
Therefore, the ``overlapping'' method can identify SLGWs from unlensed events.

\begin{figure}
\centering
\vspace{0.1em}\includegraphics[width=0.9\columnwidth]{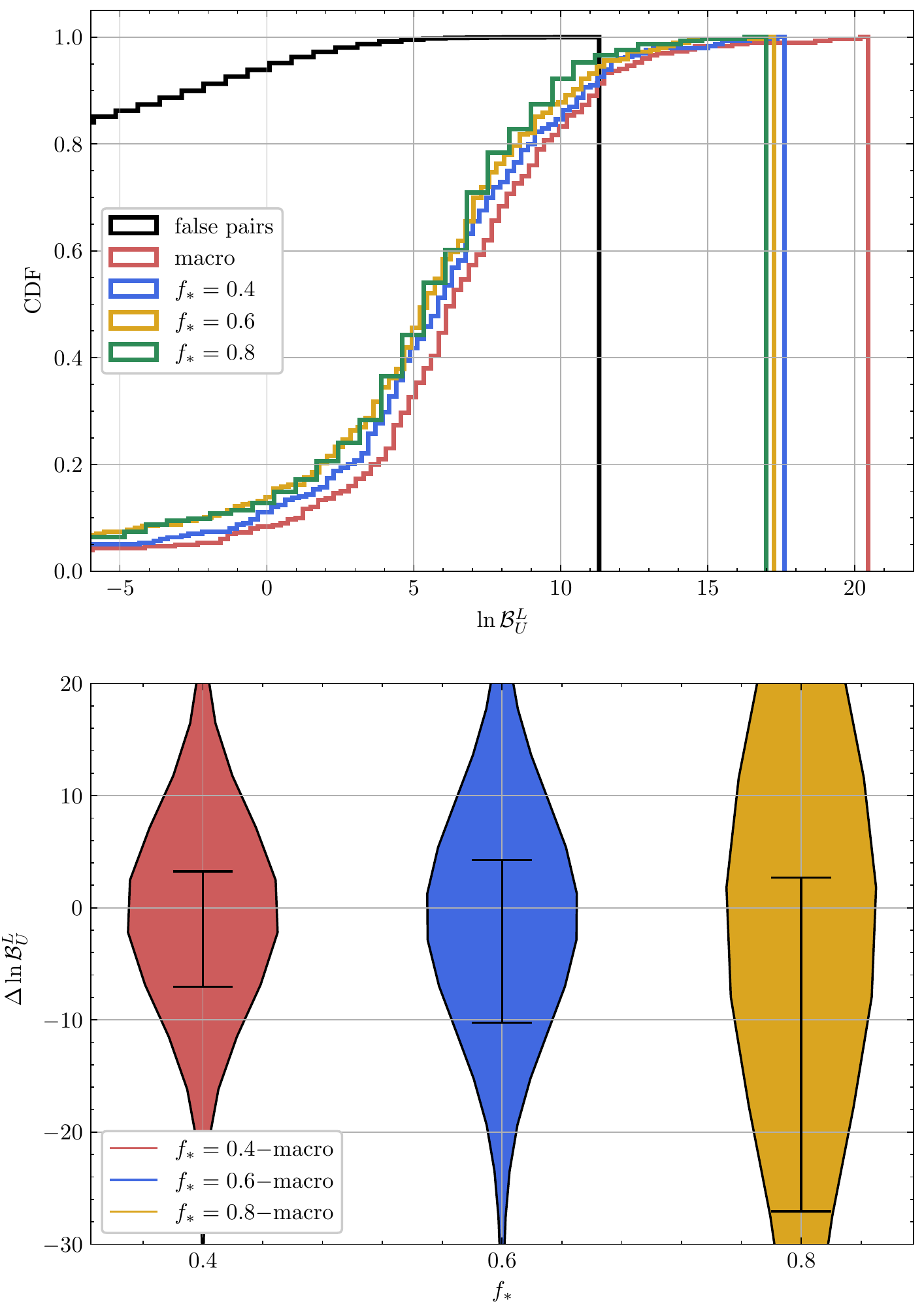}
\caption{The top panel shows the cumulative probability distribution of the Bayes factor.
The black curve stands for randomly matched unlensed events.
The red curve stands for SLGW without considering any microlensing effect.
The blue, yellow, and green curves stand for SLGWs influenced by the microlensing WO effect for different microlens densities.
The lower panel demonstrates the difference between the Bayes factor with and without the microlensing effect.
Black error bars stand for $5\%\sim95\%$ confidence intervals.
Different colors represent different microlensing densities, as shown in the legend.}  
\label{fig:bayes}
\end{figure}

The top panel of Fig.~(\ref{fig:bayes}) shows the cumulative probability distribution of the Bayes factor.
The black curve represents randomly matched pairs of unlensed events, while the red curve corresponds to SLGWs without considering any microlensing effect. 
The blue, yellow, and green curves depict the results of SLGWs influenced by microlens fields with varying densities.
For the fixed $\ln \mathcal{B}_{\mathrm{U}}^{\mathrm{L}}$ (vertical) line, the higher the value of cross point with different colored curves, the more pairs are in the low Bayes factor regime.  
It is evident that the CDF plateau can reach higher Bayes factor in the case of SLGWs than that of two randomly matched GW signals, confirming the effectiveness of the ``overlapping'' method. Besides, as the density of the microlens field increases, the maximum value of the plateau can reach decreases.

The lower panel of Fig.~(\ref{fig:bayes}) illustrates the difference in the Bayes factor with and without the microlensing WO effect.
It is observed that as the density of the microlens field increases, the Bayesian factor decreases, indicating a reduced level of match between SLGW image pairs from a statistical perspective.
This result implies that the microlensing WO effect can diminish the level of similarity between SLGW image pairs, depending on the density of the microlensing field.

\begin{figure}
\centering
\vspace{0.1em}\includegraphics[width=0.9\columnwidth]{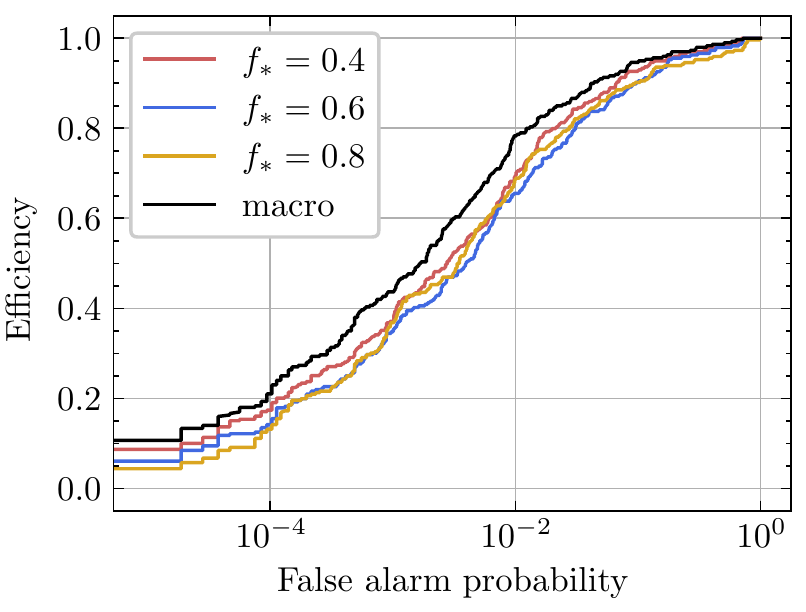}
\caption{This figure shows the influence of the microlensing WO effect on the Receiver Operating Characteristic (ROC) curves.
X-axis is the false alarm probability defined by Eq.~(\ref{eq:FAP}).
Y-axis is the detection efficiency defined by Eq.~(\ref{eq:DE}).
The black curve represents the result without considering the microlensing WO effect.
Red, blue, and yellow curves stand for the results obtained with different microlens field densities.}  
\label{fig:FPR}
\end{figure}

Fig.~(\ref{fig:FPR}) shows the impact of the microlensing WO effect on the Receiver Operating Characteristic (ROC) curves, which assess the performance of the identification method.
The x-axis is the false alarm probability per pair (${\rm FAP}_{\rm per~pair}$), defined as
\begin{equation}
\label{eq:FAP}
\mathrm{FAP}_{\rm per~pair}=\frac{N_\mathrm{unlens}(\mathcal{B_\mathrm{U}}>\mathcal{B}_0)}{N_\mathrm{unlens}(\mathrm{total})} \\,
\end{equation}
where $N_\mathrm{unlens}$ is the number of randomly matched unlensed pairs, and $\mathcal{B}_0$ is a specified Bayes factor threshold.
It describes the percentage of pairs with parameter overlaps that are greater than or equal to a threshold ($\mathcal{B}_0$) \citep{Caliskan:2022wbh}. 
The y-axis is the detection efficiency (DE), defined as
\begin{equation}
\label{eq:DE}
\mathrm{DE}=\frac{N_\mathrm{lensed}(\mathcal{B_\mathrm{L}}>\mathcal{B}_0)}{N_\mathrm{lensed}(\mathrm{total})} \\,
\end{equation}
where $N_\mathrm{lensed}$ is the number of SLGW pairs. Given a $\mathcal{B}_0$ value, one can calculate ${\rm FAP}_{\rm per~pair}$ and DE according to Eq. (\ref{eq:FAP}) and (\ref{eq:DE}). Then, one can draw this point in the Fig.~(\ref{fig:FPR}). 

The black curve is the result without considering the microlensing WO effect.
Remarkably, this curve closely aligns with the results reported in~\citep{Haris:2018vmn}, despite their focus on LIGO~\citep{2015aLIGO} and Virgo~\citep{Acernese_2014} detectors.
The consistency arises from the similarity in sky localization uncertainty between the $3$G detectors and the current detections, which ranges from $10^{-2}$ degrees to $10^4$ degrees~\citep{2017PhRvD..95f4052V}

Different colors in Fig.~(\ref{fig:FPR}) represent the results with different microlens field densities.
The findings demonstrate that the presence of the microlensing effect leads to a reduction in the detection efficiency of SLGWs.
Quantitatively, in the case of different ${\rm FAP}_{\rm per~pair}$, the DE would decrease by $0.05\sim0.15$, and the DE decreases from $\sim 0.1$ to $\sim 0.05$ at $\mathrm{FAP}_{\rm per~pair}=10^{-5}$.
This result means that the microlensing field introduces the possibility of missing certain SLGW events across all ${\rm FAP}_{\rm per~pair}$ scenarios, and nearly half of the candidates being missed at $\mathrm{FAP}_{\rm per~pair}=10^{-5}$.
Therefore, people need to pay more attention to the microlensing WO effect for SLGWs.

\begin{figure*}
\centering
\vspace{0.1em}\includegraphics[width=\textwidth]{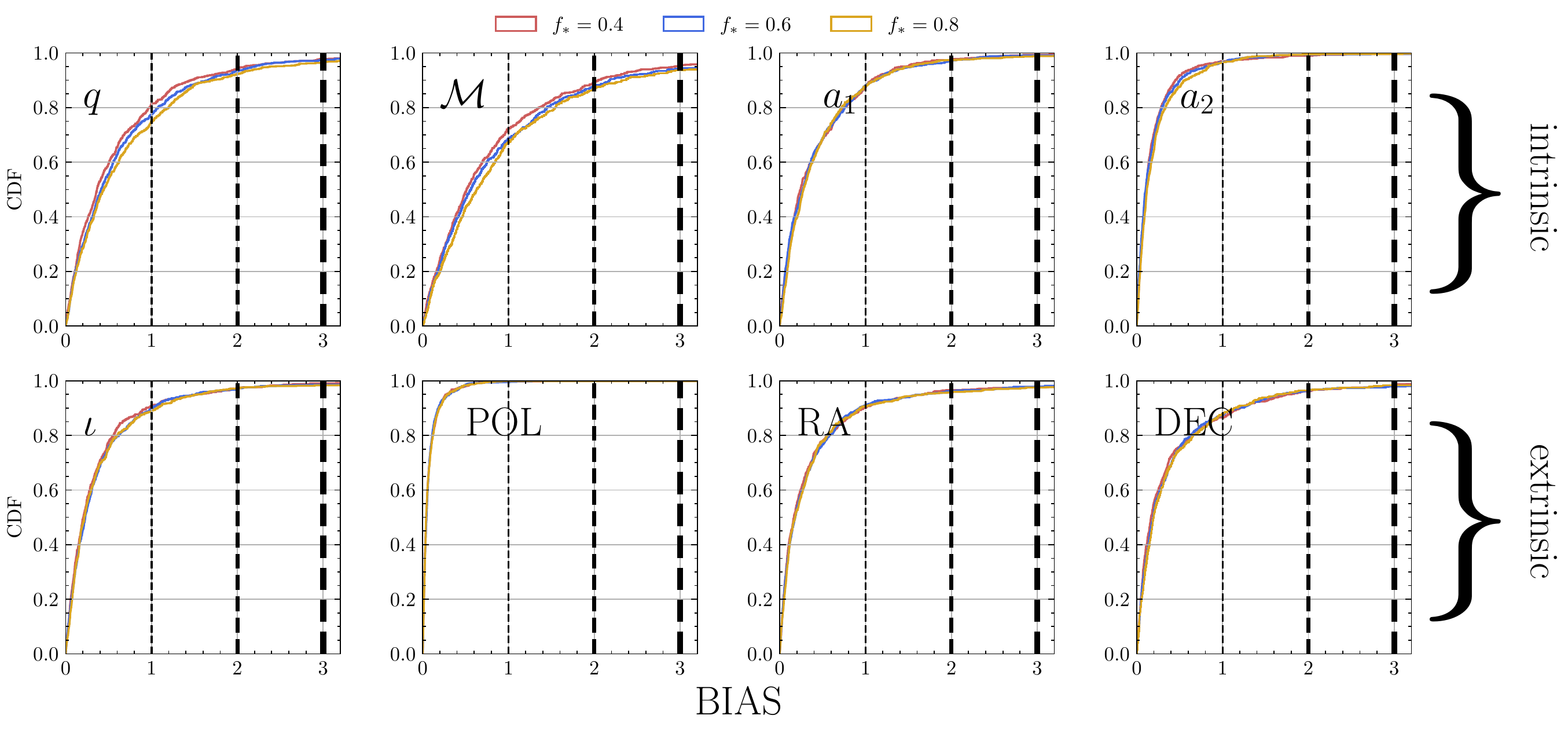}
\caption{This figure illustrates the impact of microlensing on the estimation of SLGW parameters.
The x-axis represents the relative bias, defined by Eq.~(\ref{eq:rel_bias}).
The y-axis displays the cumulative probability distribution function.
The first row shows the results for intrinsic parameters (mass-ratio $q$, chirp mass $\mathcal{M}$, and dimensionless spins $a_1$ and $a_2$).
The second row shows the results for extrinsic parameters (inclination angle $\iota$, polarization angle POL, right ascension RA, and declination DEC).
Distinct colors correspond to different microlensing field densities, as indicated in the legend.}  
\label{fig:Rel_bias}
\end{figure*}

Fig.~(\ref{fig:Rel_bias}) illustrates the microlensing-induced bias on the SLGW parameters.
The x-axis represents the relative bias, defined as
\begin{equation}
\label{eq:rel_bias}
\mathrm{BIAS} = \frac{|\bar{x}_{\text {micro }}-\bar{x}_{\text {macro }}|}{\sqrt{\sigma^2\left(x_{\text {macro }}\right)}} \,
\end{equation}
where $\bar{x}_{\text {micro}}$/$\bar{x}_{\text {macro}}$ denotes the mean value of the parameter with/without microlensing influence, and $\sqrt{\sigma^2\left(x_{\text {macro }}\right)}$ is the standard deviation of the posterior distribution without microlensing effect.
Here, $x$ represents the GW parameters listed in the figure.
The y-axis is the cumulative probability distribution function of the quantity defined in Eq. (\ref{eq:rel_bias}).
The first row presents the results for intrinsic parameters, including mass-ratio $q$, chirp mass $\mathcal{M}$, and dimensionless spins $a_1$ and $a_2$.
The second row shows the results for extrinsic parameters, including inclination angle $\iota$, polarization angle POL, right ascension RA, and declination DEC.
The different colors correspond to different microlens field densities, as indicated in the legend.

It is observed that the microlensing bias is more pronounced for intrinsic parameters compared to extrinsic parameters. 
Particularly, for the mass ratio (chirp mass), more than $20\%$($30\%$) of events exhibit biases greater than $1\sigma(x_\mathrm{macro})$, and more than $2\%$($5\%$) of events display biases greater than $3\sigma(x_\mathrm{macro})$.
We consider these $3\sigma$ biased events as purely biased.
This phenomenon arises from the dominance of waveform in these two parameters, which are tightly constrained by the $3$G detectors.

Regarding the impact on other parameters, we observed $1\sigma$ biases across different levels, while $3\sigma$ biases were found to be rare.
Additionally, in line with our understanding, it is evident that a higher density of the microlens field leads to an increased number of biased events. 
For convenience, we have summarized the key findings in Table~\ref{ta:BBHPara_bias} and showing a parameter estimation result in Appendix~(\ref{ap:PER}).

\begin{table*}
  \centering
  \caption{\label{ta:BBHPara_bias} The percentage of events exhibiting biases greater than $1$, $2$ and $3~\sigma$ is presented for various parameters and microlensing densities ($f_*$).} 
   \begin{tabular}{l|cccccccc} 
    \hline
    \diagbox{$f_*$}{$\#$~$\%$ ($1/2/3$~$\sigma$)}{} & $q$ & $\mathcal{M}$ & $a_1$ & $a_2$ & $\iota$ & POL & RA & DEC \\ 
    \hline
    $0.4$ & $19$/$6$/$2$ & $28$/$11$/$4$ & $13$/$2$/$0.6$ & $3$/$1$/$0.3$ & $9$/$3$/$0.9$ & $0.1$/$0$/$0$ & $10$/$3$/$2$ & $13$/$4$/$2$ \\
    \hline
    $0.6$ & $23$/$7$/$3$ & $31$/$12$/$6$ & $12$/$3$/$1$ & $3$/$0.7$/$0.4$ & $10$/$3$/$1$ & $0.4$/$0$/$0$ & $9$/$4$/$2$ & $12$/$4$/$2$ \\
    \hline
    $0.8$ & $25$/$8$/$3$ & $33$/$13$/$6$ & $12$/$3$/$1$ & $3$/$0.5$/$0.4$ & $11$/$3$/$2$ & $0$/$0$/$0$ & $9$/$4$/$2$ & $12$/$3$/$2$ \\
    \hline
  \end{tabular}
\end{table*}

\section{Summary and Discussion}
\label{sec:sum_dis}
The evaluation of the ``overlapping'' level between two signals is a commonly used method for identifying SLGWs. 
However, the presence of microlensing fields within the SLGWs can introduce frequency-dependent effects on the SLGW waveforms, potentially leading to biases in parameter estimation and affecting the identification of SLGW pairs. 
The objective of this paper is to provide a quantitative assessment of the impact of microlensing on SLGW identification and parameter estimation.

Fig.~(\ref{fig:bayes}) and Fig.~(\ref{fig:FPR}) illustrate the influence of the microlensing WO effect on the Bayes factor for ``overlapping'' degree and the efficiency of SLGW identification. 
The results reveal that the presence of a microlensing field can lead to a decrease in the Bayes factor and a reduction in the identification efficiency of SLGWs. 
Notably, at various false alarm probabilities, the detection efficiency get reducedy by $5\%$ to $50\%$. At a false alarm probability of $10^{-5}$, the detection efficiency decreases from approximately $0.1$ to $0.05$. 
This implies that nearly half of the SLGW candidates could be missed at a false alarm probability of $10^{-5}$. 
Therefore, it is crucial to pay closer attention to the microlensing WO effect.

Fig.~(\ref{fig:Rel_bias}) investigates the impact of the microlensing WO effect on the parameter estimation of SLGWs under different microlensing density scenarios. 
The results highlight two main conclusions. 
First, the microlensing WO effect introduces a noticeable bias in intrinsic parameters, particularly for $q$ and $\mathcal{M}$, but has little influence on extrinsic parameters. 
Thus, the SLGW identification loss depicted in Fig.~(\ref{fig:FPR}) is primarily attributed to the influence on intrinsic parameters. 
Second, a higher density of microlensing fields leads to a greater number of biased events. 
In sumarry, considering all parameter biases, approximately $30\%$ of events exhibit a $1\sigma$ bias, approximately $12\%$ exhibit a $2\sigma$ bias, and approximately $5\%$ exhibit a $3\sigma$ bias.

This paper investigates the influence of the microlensing WO effect on SLGW identification and parameter estimation, offering a more accurate theoretical prediction for future SLGW identification. 
The results emphasize the significance of considering the microlensing WO effect in the analysis of SLGW events and finding effective strategies to mitigate the biases introduced by microlenses.

\section*{Acknowledgments} 
This work is supported in part by the National Key R\&D Program of China No. 2021YFC2203001. 

\section*{Data Availability}
The data underlying this article will be shared on reasonable request to the corresponding author.

\section*{Annotation}
As our article approached its final stages, a comparable paper surfaced on arXiv~\citep{Mishra:2023ddt}, presenting congruent conclusions with some aspects of our own research.

\appendix

\section{Parameter estimation Result}
\label{ap:PER}
In this appendix, we shows the parameter estimation result for the SLGW event displayed in Fig.~(\ref{fig:waveform}).
The red and blue colors representing the results without and with the microlensing effect, respectively.
The black curves indicate the injected values for various parameters.
One can see an evident shift in parameter estimation (particularly for the $2$D distribution of $q$ and $\mathcal{M}$) from the injected value towards a biased value in the microlensing scenario.
It is important to emphasize that this SLGW event displays a noticeable parameter bias at a $3\sigma$ confidence level (Eq.~(\ref{eq:rel_bias})).

\begin{figure*}
\centering
\vspace{0.1em}\includegraphics[width=\textwidth]{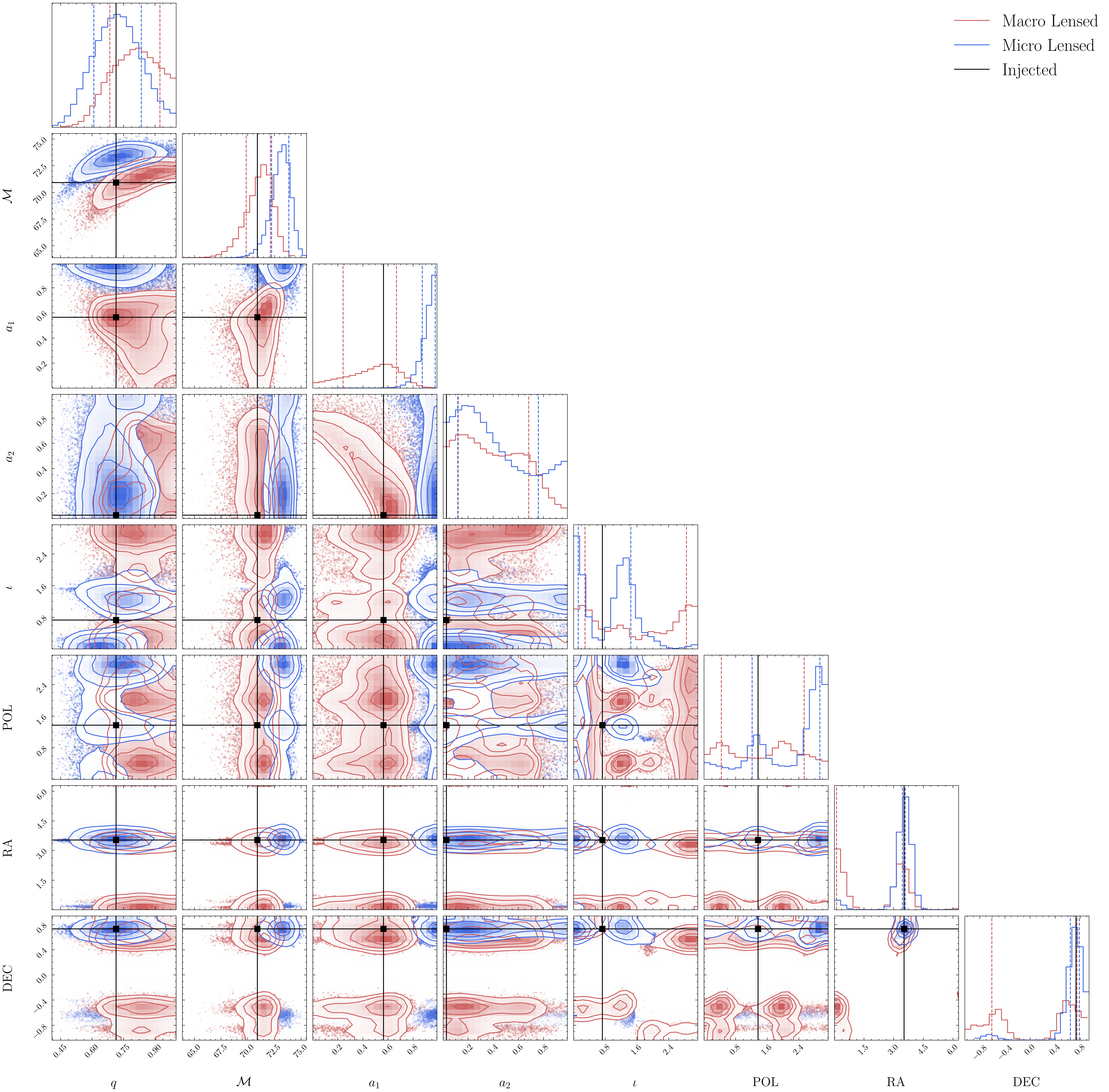}
\caption{
This figure depicts the parameter estimation results for SLGW event displayed in Fig.~(\ref{fig:waveform}), employing the \texttt{Bilby} package.
The red and blue colors correspond to the results of SLGW without and with the microlensing effect, respectively. 
The black curves represent the injected values for various parameters.
}  
\label{fig:PER}
\end{figure*}



\bibliographystyle{mnras}
\bibliography{ref} 




\appendix


\bsp	
\label{lastpage}
\end{document}